\documentstyle[mprocl]{article}

\begin{document}

\title{Angular Noise in Gravitational Wave Detectors}

\author{Gabriela Gonz\'alez}

\address{
Center for Gravitational Physics and Geometry\\ 
The Pennsylvania State University \\
104 Davey Laboratory, University Park, PA 16802, USA}

\maketitle

\abstracts{Angular fluctuations of suspended mirrors in
gravitational wave interferometers are a source of noise both for the
locking and the operation of the detectors. We describe here some of
the sources of these fluctuations and methods for the estimation of
their order of magnitude.}

In interferometric gravitational wave detectors such as LIGO, the
mirrors are suspended on fine wires. In designs for advanced
detectors, the pendulums where the mirrors hang have multiple stages
with both horizontal and vertical soft spring constants. The pendulum
frequencies (around or below 1 Hz) are much lower than the frequency
with optimal sensitivity for gravitational waves (around 100 Hz).  The
suspended configuration, which is necessary for many reasons (free
masses, seismic isolation and immunity to thermal noise), has the
disadvantage of allowing large pendular motions that need to be
controlled to keep the detectors in a linear regime. Much attention
has been paid to the horizontal motion of the masses, since they are
directly observed by the interferometer and thus limit the sensitivity
to gravitational waves.

However, the {\em angular} motion of the mirrors can also be an
important source of sensitivity degradation, especially in long
baseline optical cavities\cite{ApplOpt}.  Requirements for the
acceptable amount of angular noise come in two flavors: ({\em i}) How
much before we cannot lock easily the interferometer?  ({\em ii}) How
much can be tolerated while in operation? For the LIGO detectors now
in construction, answers are ({\em i}) 0.5 mrad rms and ({\em
ii})$10^{-8}$ rad rms, and contributions less than 5\% of shot noise
\cite{ASCReqts}. For advanced interferometers, the requirements may be
up to 10 times tighter, which is very difficult to achieve: this
highlights the need for their accurate estimation.

The identified sources of angular noise are straight seismic
excitation (depends on pendulum mechanics); seismic excitation through
mechanical cross-coupling; electronic excitation due to
sensing/control noise; electronic excitation through actuators. 
cross couplings; and the always unavoidable thermal noise.

Seismic noise produces angular noise through straightforward pendulum
dynamics. In LIGO these calculations predict motions smaller than the
required tolerance. However, due to mechanical asymmetries, it is
likely that a larger angular motion is produced by cross couplings of
other degrees of freedom. This problem will be compounded in multiple
pendulums, and needs to be carefully studied. We have an experimental
program in Penn State to such effect \cite{AmaldiGG}.

The control of the masses done to damp the pendular motions is usually
applied as several forces on different points of the mirror that add
up to the required forces and torques derived in a feedback mechanism
from sensors at the same points where the actuation is done. The sensed
signals can be decomposed in the natural suspension modes, with a
different feedback mechanism for each mode (``modal damping''), as is
done on LIGO suspensions, or they can be sensed and acted upon
independently, as long as the system is not overdetermined
(``point-to-point damping''), as is done in GEO600 suspensions. Both of
these systems, however, suffer the risk of introducing excess forces
due to the imperfect and unbalanced sensing or actuation. The sensor
noise ($\approx 10^{-9} {\rm rad/\sqrt{Hz}}$) introduces excess
angular motion just below the tolerances for the LIGO detector, but
the same sensors will be too noisy for advanced LIGO detectors. This
is however, a problem only relevant for locking acquisition: when in
operation, interferometric (``wavefront'') sensors can be used whose
noise is negligible compared with tolerances. The limiting noise in
the actuators is severely constrained by the maximum horizontal motion
they can produce, so there are no additional constrained posed by the
angular noise they may produce. However, the angular noise introduced
by unbalanced forces is a serious problem in the LIGO detectors that
needs a careful tuning of the actuators, taking into account the
pendulum dynamics \cite{SuspTuning}. 

The power spectral density of the thermal noise, or brownian motion,
of the mirrors is determined by the dissipation mechanisms in the
suspension, which in the ideal situation are due to material
properties of the suspensions wires. A calculation of the angular
thermal motion can be done using the Fluctuation-Dissipation theorem,
but the needed calculation is the observation of the angular motion by
the interferometer: this depends on the angles between the axes of the
optical cavities and the suspension axes. This calculation has been
done for LIGO suspensions by the author \cite{CQG}, and shows that the
angular thermal noise will not limit the detector sensitivity. However,
this calculation needs to be done for advanced detectors, where
angular noise may represent a significant contributor.

In conclusion, we suggest that angular noise of suspended mirrors does
not limit the sensitivity of LIGO detectors, but only when extreme
care is taken care with control cross couplings. The safety margin is
less than an order of magnitude for some of the sources, which
suggests that angular noise needs to be carefully studied in advanced
LIGO configurations.

\end{document}